\documentclass[final,3p,times,twocolumn,sort&compress]{elsarticle}

\usepackage{graphics}

\usepackage{amssymb}
\usepackage{color}
 \usepackage{amsthm}
\usepackage{amsmath}
\usepackage{tabularx}

 \journal{Acta Materiala}

\begin{document}

\begin{frontmatter}


\title{Fiber Be \tnoteref{}}
 \author[cemes]{Mompiou, F.\corref{cor1}} 
 \ead{mompiou@cemes.fr}
 
 \author[cemes]{Legros, M.}
 \ead{legros@cemes.fr}

\author[KIT]{Ensslen, C.}
 \ead{charlotte.ensslen@kit.edu}
 
 \author[KIT]{Kraft, O.}
 \ead{oliver.kraft@kit.edu}

 \cortext[cor1]{Corresponding Author}
 
\address[cemes]{CEMES-CNRS and Universit\'e de Toulouse, 29, rue J. Marvig, 31055 Toulouse, France}
\address[KIT]{Karslruhe Institute of Technology, Hermann-von-Helmholtz-Platz 1, 76344 Eggenstein-Leopoldshafen, Germany}


\title{In situ TEM study of twin boundary migration in sub-micron Be fibers}

\begin{abstract}
Deformation twinning in hexagonal crystals is often considered as a way to palliate the lack of independent slip systems. This mechanism might be either exacerbated or shut down in small-scale crystals whose mechanical behavior can significantly deviate from bulk materials.  Here, we show that sub-micron beryllium fibers initially free of dislocation and tensile tested in-situ in a transmission electron microscope (TEM) deform by a $\{ 10\bar{1}2 \}$ $\langle 10\bar{1}1 \rangle$ twin thickening.  The propagation speed of the twin boundary seems to be entirely controlled by the nucleation of twinning dislocations directly from the surface. The shear produced is in agreement with the repeated lateral motion of twinning dislocations. We demonstrate that the activation volume ($V$) associated with the twin boundary propagation can be retrieved from the measure of the twin boundary speed as the stress decreases as in a classical relaxation mechanical test. The value of $V \approx 8.3 \pm 3.3 \times 10^{-29}m^3$ is comparable to the value expected from surface nucleation. 

\textbf{This text is a revised version of the article published in Acta Materialia, 96 (2015) 57-65 (doi:10.1016/j.actamat.2015.06.016)}
\end{abstract}


\end{frontmatter}


\section{Introduction}
\label{intro}

Hexagonal close packed (hcp) metals are known for their particular mechanical properties such their  high strength to weight ratio.
In such anisotropic metals, the mechanical response strongly depends on the solicitation and texture. For instance, Hall-Petch effects appear to be dependent on the orientation and Peierls friction stress on the different slip systems \cite{Sharon2014}. Deformation twinning is often easier than the activation of non basal dislocation slip, which explains its importance in the deformation of many hexagonal metals \cite{Christian1995} as it may be a way to enhance the ductility \cite{Yoo1981}.

However, twinning also induces a strong strain hardening as twins act as barrier to dislocation motion and to further twin propagation \cite{Gutierrez2012, Kalidindi2003}.
In the recent years, the possibility of  a fine tuning of the mechanical properties by incorporating twins in the microstructure, such as nanotwinned copper \cite{Lu2009} or twinning induced plasticity (TWIP) steels \cite{Bouaziz2011},  have been extensively used. In finely twinned copper for instance, the decrease of the twin separation distance down to few nanometers have led to a large increase of the strength while maintaining some ductility. However, a clear picture of the impact of twinning on the mechanical properties of metals is still lacking, especially at small scales \cite{Zhu2012, Liu2014}, mainly because several competing mechanisms can be at play: twin nucleation,  propagation and thickening, interaction between twins or with dislocations \cite{Chassagne2011}, de-twinning \cite{Morrow2014} etc.  
The kinetics of plastic deformation can be investigated by measuring the sensitivity of the stress to the strain rate and to the temperature to determine the activation parameters of rate controlling mechanisms. The activation volume $V$ which is inversely proportional to the strain rate sensitivity is a measure of the volume of matter involved in the thermally activated process \cite{Caillard2003}. Such parameters (with the activation energy) are often used to discriminate between these various mechanisms but are often complex to obtain, especially during an in situ TEM experiment. In nanotwinned copper, it has been postulated from the low value of the activation volume (few $b^3$, compared to value of several hundreds of $b^3$ for typical mechanisms in coarse grained) that twin transmission is the rate controlling deformation mechanism \cite{Lu2009b}. Moreover, because lattice dislocations can easily dissociate and eventually glide inside a coherent twin \cite{Chassagne2011}, contrary to grain boundaries where they usually decompose \cite{Mompiou2012, Kwiecinski1991a}, ductility is enhanced compared to conventional nanocrystalline (nc) metals.

Beryllium deforms by dislocation slip on the basal and prismatic planes \cite{Poirier1967,Regnier1970}, but prismatic glide exhibits a strong anomalous increase of the yield stress with temperature close to the room temperature, which can be ascribed to strong friction stresses \cite{Beuers1987, Couret1989}.
Twinning through gliding dislocations mechanisms have been  proposed early in hcp metals \cite{Cottrell1951,Thompson1952,Westlake1961, Serra1988}. 
For the $\{ 10\bar{1}2 \}$ $\langle 10\bar{1}1 \rangle$ twin, which is the activated system in Be, it involves dislocations of Burgers vector $\approx 0.12 [10\bar{1}1]$ associated with a step which height is 2 times the $\{10\bar{1}2\}$ interplanar spacing (zonal twinning dislocation) \cite{Christian1995}. In a general context of interfacial defects, such dislocations with a step character have been coined disconnection by Hirth and Pond \cite{Hirth1996}.
If twinning dislocations are well known, their propagation and especially their nucleation mechanism still remains partly understood. 
Both homogeneous or heterogeneous nucleation and propagation can be considered \cite{Christian1995}. 
The first case is thought to arise when a combination of very high stress and very low surface and strain energy is satisfied. 
Although unlikely in conventional bulk crystals, this process may happen in dislocation free whiskers as suggested by early observations of Price in Zn \cite{Price1960}, recent ones by Roos in Au \cite{Roos2014} or  in small Mg pillars \cite{Yu2012a}. 
Twin thickening by the spontaneous nucleation of twinning dislocations has also been envisaged to account for the observation of growing twins in Cu-Ge alloys without evidence of source operation or dislocation emission from grain boundaries \cite{Narayan2008}.
Such high stress concentrations can also be found near crack tip where twin nucleation can be observed \cite{Shu2013} or from grain boundaries and triple junctions \cite{Sennour2007}, especially in nanograins where partial dislocation nucleation is thought to be less energetically costly than perfect dislocations \cite{Chen2003, VanSwygenhoven2006}.
From a theoretical point of view, it is now well established that dislocation nucleation from free surfaces can be facilitated by surface imperfections \cite{Godet2004, Guenole2013}. 
In the case of twinning dislocations, it has been suggested that the nucleation of a second twinning dislocation in an adjacent plane can be stimulated by the presence of the stacking fault of the first one \cite{Yu2010,Yu2012a}.
Recent molecular dynamics simulations in Mg also indicate that twin thickened from the formation of twinning dislocations presumably from the surface \cite{Li2009b}.

\begin{figure*}
\centering
	\includegraphics[width=0.6\textwidth]{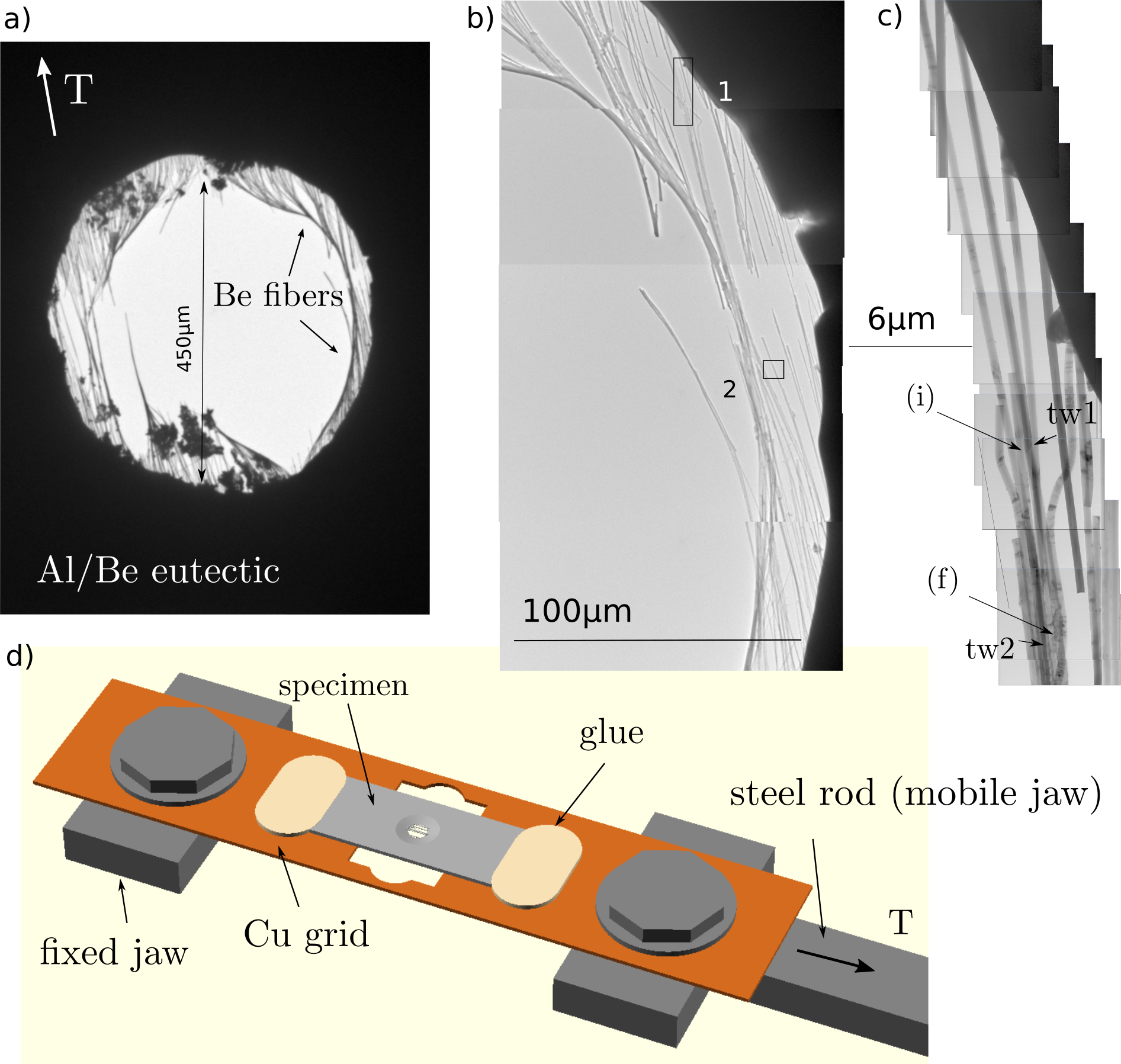}
	\caption{a) A selectively etched Al/Be eutectic forming a hole containing freestanding electron transparent Be fibers. b) is a zoom on the right side of the hole where fibers are attached from both sides to the rest of the eutectic. c) is a zoom in the region 1 shown in b) where the observations  were performed. d) is a schematics of the straining device. The straining direction $T$ is along the fiber direction.}
\label{fig-gage}
\end{figure*}
Heterogeneous nucleation through the operation of a twinning dislocation source composed of a sessile dislocation intersecting the twin boundary, i.e. the pole mechanism has been proposed early in hcp metals \cite{Thompson1952}. 
More recently, Serra and Bacon have found that the decomposition of a lattice dislocation into a $\{ 10\bar{1}2 \}$ twin creates a sessile disconnection that eventually acts as a source \cite{Serra1996} which follows the twin boundary as it migrates. 
Recent mechanical modeling tends to indicate however that such decomposition of lattice dislocation into twinning dislocations (so-called slip-assisted twinning) is insufficient to account for the mechanical properties of hcp metals and so that direct twinning nucleation should also operate \cite{Capolungo2009}.	
Thus, the twin thickening under a shear stress parallel to the twin boundary occurs by the repeated motion of twinning dislocations. 
A similar situation has also been found for grain boundaries, where the migration coupled to stress is due to the motion of disconnections \cite{Khater2012, Rajabzadeh2013, Rajabzadeh2013a}. 
This  phenomenon which has been observed either in bicrystals \cite{Molodov2007,Gorkaya2009} and polycrystals \cite{Rupert2009, Mompiou2009, Mompiou2011} plays an important role in the mechanical properties of nc metals by promoting grain growth.

In this paper, we report the observation of a single $\{ 10\bar{1}2 \}$ $\langle 10\bar{1}1 \rangle$ twin boundary migration in a sub-micron beryllium fiber deformed in tension in-situ in a transmission electron microscope (TEM). After presenting the microstructure and the experimental set-up, we show that the twin boundary did not propagate by the operation of a pole mechanism, but more presumably by the nucleation from the free surface. 
By measuring the twin boundary speed as the stress relaxed, we were able to retrieve the activation volume associated with this individual mechanism. We then discuss both the nucleation and motion of twin dislocation mechanisms to interpret the obtained value.

\section{Experimental}
\label{experimental}
Freestanding beryllium fibers where obtained from a directionally Al-2.4 at.$\%$Be eutectic alloy with the same approach described in \cite{Mompiou2012b}.
3 by 1mm samples were first cut in a bulk ingot by an electrodischarging machine with the long direction parallel to the fiber length. They were then mechanically polished and etched by the Struers A2 electrolytic solution at -10$^\circ$C. 
This leads to the fast etching of the aluminium matrix which freed the Be fibers in a circular etched region as shown in figure \ref{fig-gage}a.
In consequence, the fibers are freestanding in the middle of a TEM hole. 
Around the hole, the rest of the specimen is composed of the eutectic alloy.
The figure \ref{fig-gage} shows the geometry of the TEM hole with the fibers. 
The complete hole (figure \ref{fig-gage}a) is 450 $\mu$m large and no electron transparent zones can be observed on its borders meaning the eutectic alloy has been etched only in its center. 
The hole itself is composed of several tens of fibers of approximately the same diameter. 
Most of the fibers have only one end attached to the rest of the matrix.
The area where we concentrate our investigation is located on the right side of the hole close to the edge and contains few fibers attached at both ends (figure \ref{fig-gage}).
A GATAN straining holder was used at room temperature to deform the whole specimen containing the fibers.
During this mechanical test, the displacement is imposed via a long steel rod to the sample fixed on a copper grid (see figure \ref{fig-gage}d). 
The copper grid is attached to the jaws by screws. 
The TEM sample itself is glued on the copper grid. 
In-situ observations were performed on a JEOL2010 operated at 200kV. Video sequences were recorded on a DVD by a MEGAVIEW III camera at 25fps.
Tensile tests of individual fibers were performed in a Nova 200 scanning electron microscope/focused ion beam (SEM/FIB) equipped with a piezo-actuated platform and Femtotools load cells.  Strain is calculated using a Matlab routine \cite{Eberl2006}. For more details about individual wire testing, one can refer to \cite{Gianola2011a,Richter2009,Johanns2012}.

\section{Results}
\begin{figure}
\centering
	\includegraphics[width=0.35\textwidth]{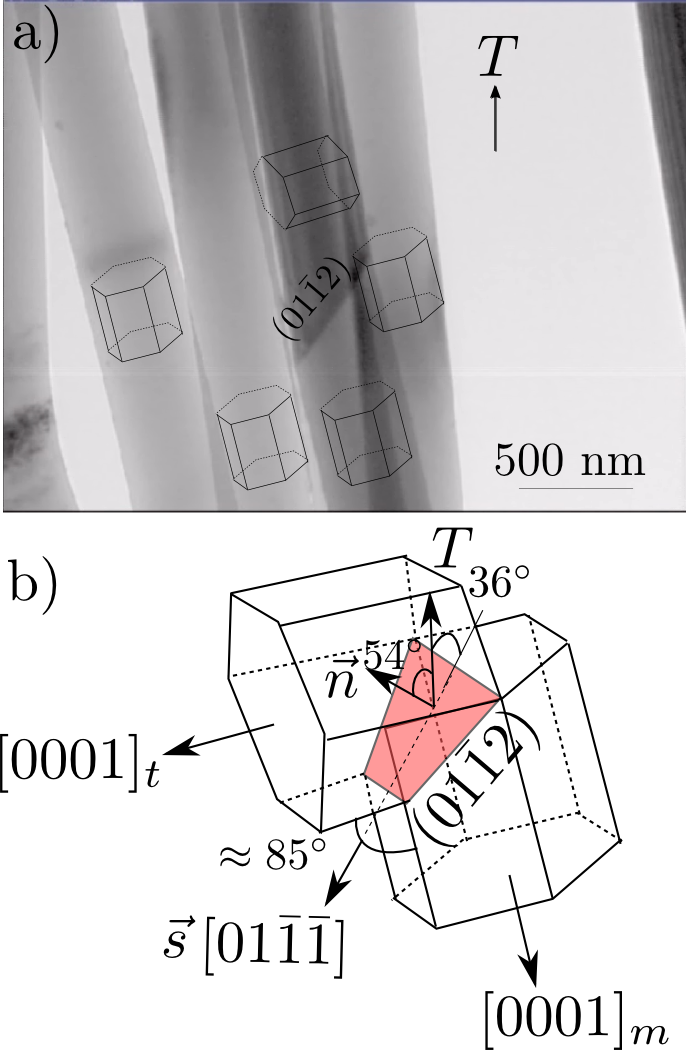}
	\caption{a) Most of the fibers deformed along the straining axis $T$ having their long direction parallel to their $\langle c \rangle$-axis are hard to deform while the $(01\bar{1}2)$ twin is favourably oriented (b).}
\label{fig-twin-geo}
\end{figure}
\begin{figure*}
\centering
	\includegraphics[width=0.65\textwidth]{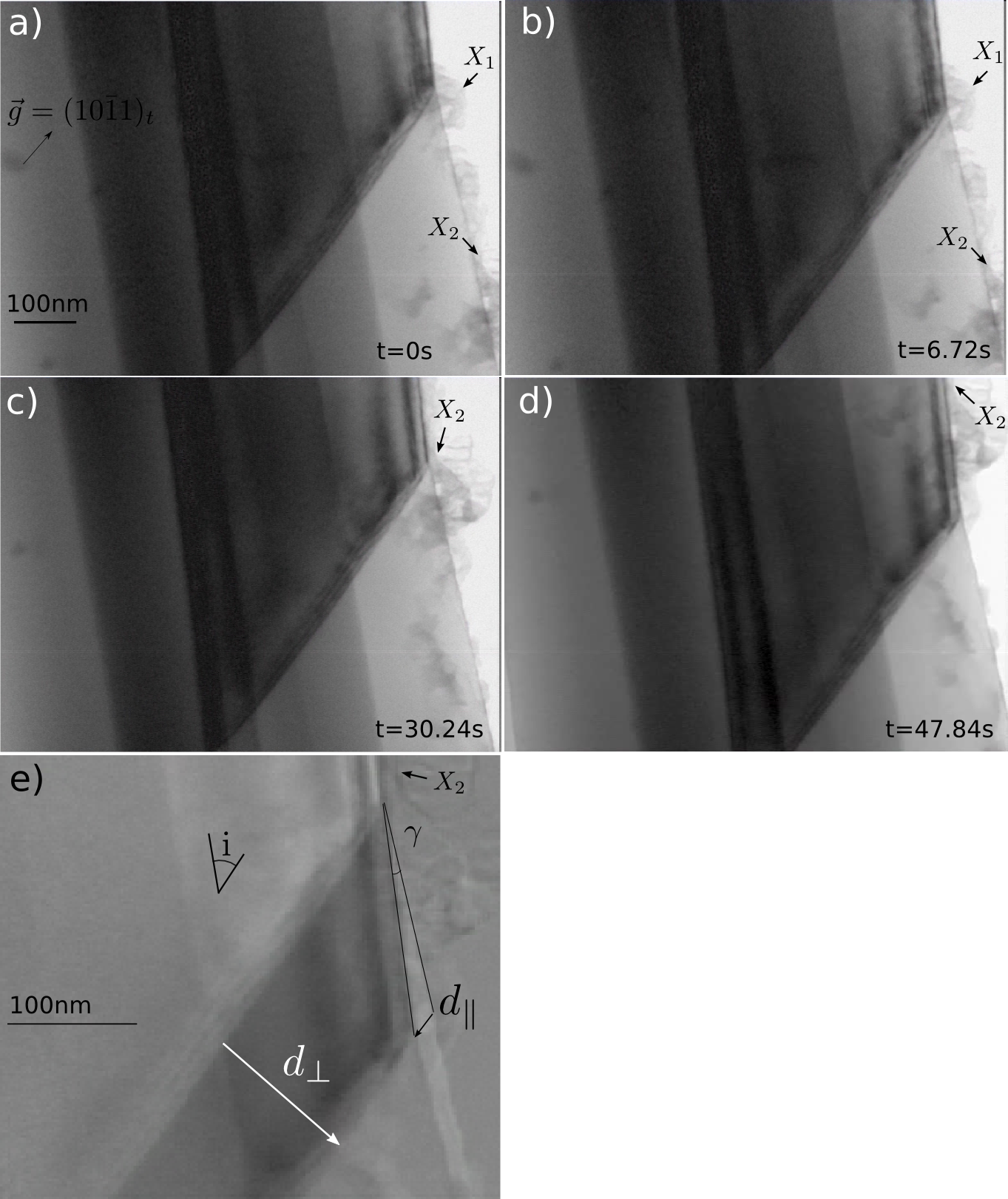}
	\caption{a) to d) show the migration of a $(01\bar{1}2)$ twin boundary under stress. $X_1$ and $X_2$ are fiducial markers. e) is the image difference d)-c) when $X_2$ position is set as a reference. This allows the measure of both the migration distance ($d_\perp$) and the shear displacement ($d_\|$).}
\label{fig-twin-migration}
\end{figure*}
Most of the fibers in the area shown in figure \ref{fig-gage}c are about 200 $\mu$m long. 
End-on observations of broken fibers by SEM show that the fibers have a circular cross section. 
The diameter of the fibers is almost constant over the whole fiber length. 
The majority of the fibers have a diameter of approximately 400 nm but fibers with  diameter around 250 nm were also observed.
The fibers are usually straight and no dislocations can be observed.
Fibers which are detached at one end can be bent, and a high dislocation density can be observed in the bent area.
This indicates that plastic deformation until fracture has occurred prior the experiments in a very limited area, similarly as observed in aluminium fibers \cite{Mompiou2012b}.
More details on plastic deformation by dislocation slip will be given in a forthcoming article.
Few $\{ 10\bar{1}2 \}$ twin boundaries can also be observed. 
They are indicated in figure \ref{fig-gage}c by the label $tw1$ and $tw2$. 
Another twin can be observed in the area 2 where fibers are already broken (figure \ref{fig-gage}b). 
Lots of fibers stack together forming a large loose tangle in figure \ref{fig-gage}b.
Figure \ref{fig-twin-geo} shows a closer view of the twin boundary $tw1$.
 Electron diffraction analysis has shown that most of the fibers have their $\langle c \rangle$ axis parallel to the fiber length  while the twinned area present a $\langle c \rangle$ axis almost perpendicular (figure \ref{fig-twin-geo}a), so that the $(01\bar{1}2)$ twin plane is making an angle of about $54^\circ$ with the straining axis $T$ (figure \ref{fig-twin-geo}b). 
In such configuration, the Schmid factor, $s_f=\cos(\vec{n},\vec{T}) \cos(\vec{s},\vec{T})\approx 0.47$, with $\vec{n}$ the normal to $(01\bar{1}2)$  and $\vec{b}$ is the shear direction, i.e. parallel to $[01\bar{1}\bar{1}]$, is close to the maximum value. Such configuration then favours the growth of the twin in tension.

\subsection{Twin boundary migration}

Twin boundary migration under stress has been observed after strain increment until the stress relaxed up to a critical value where the twin boundary remained immobile. 
After several increments, the twin boundary $tw1$ seen at its initial position (i) in a 400 nm large fiber (figure \ref{fig-gage}c) has moved to its final position (f), which represents a distance of about 8.5 $\mu$m, before being stopped definitely. 
A second twin boundary $tw2$ located on an adjacent fiber (diameter around 400 nm)  in the lower part of area 1 (figure \ref{fig-gage}) started then to move. 
The motion of this twin boundary has been followed over 5 $\mu$m.
Figure \ref{fig-twin-migration} shows bright field images, taken under the diffraction vector $\vec{g}=(10\bar{1}1)_t$ (the subscript $t$ refers to the twinned area), extracted from a video sequence. 
They show the motion of the twin boundary $tw1$ over a short distance after a strain increment. 
The motion of the twin boundary can be easily followed by remarking fixed points located at the fiber surface and noted $X_1$ and $X_2$ (for the corresponding video see the supplementary materials). 
The twin boundary motion appears to be smooth with an average migration speed around 10 nm/s.
Figure \ref{fig-twin-migration}e is the image difference between figure \ref{fig-twin-migration}d and figure \ref{fig-twin-migration}c when the marker $X_2$ is set as a reference (see the uniform grey contrast of $X_2$).
It can then be clearly seen that the  twin boundary migration has produced a displacement of the lower part of the fiber of a distance $d_\|$. 
This displacement linearly increases as the twin boundary migrates,  as evidenced by the angle $\gamma$ formed by the fiber sides on both sides of the twin boundary (figure \ref{fig-twin-migration}e). 
As a consequence, the fiber is deformed by a shear strain also called coupling factor ($\beta$) for grain boundaries by Cahn et al. \cite{Cahn2004}.

\begin{figure}
\centering
	\includegraphics[width=0.45\textwidth]{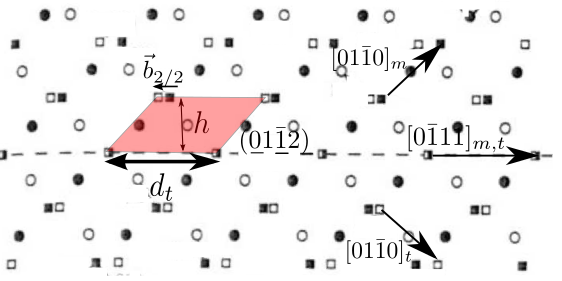}
	\caption{Projection of the dichromatic pattern of the twin along the $[2\bar{1}\bar{1}0]$ direction. Black and white symbols corresponds to the twin and matrix respectively. Circle and square symbols indicate lattice positions in two adjacent planes along the $[01\bar{1}2]$ direction. The twinning dislocation has a Burgers vectors $\vec{b}_{twin}=\vec{b}_{2/2}$ associated with a step $h$. The shaded red area corresponds to the area swept by the twinning dislocation while moving along the twin plane during the twin boundary migration.}
\label{fig-dichro}
\end{figure}

Since the twin plane is inclined only $6^\circ$ from the electron beam direction in figure \ref{fig-twin-migration}, the shear strain can be directly obtained by measuring the inclination of the twin plane with respect to the fiber direction (angle $i$) and the angle $\gamma$:
\begin{equation}
\beta=\frac{d_\|}{d_\perp}=\frac{\sin \gamma}{\sin i \sin(i-\gamma)}
\end{equation}
Taking $i=42 \pm 1^\circ $ and $\gamma=5.4 \pm 0.4 ^\circ$ leads to $\beta=0.237\pm 0.03$.
This value is very close to the value expected as explained below.
The figure \ref{fig-dichro} shows the $[1\bar{2}10]$ projection of the dichromatic pattern of the twin. 
Black symbols correspond to the matrix ($m$) and white ones to the twinned crystal ($t$) while opened and close symbols indicate the lattice positions in two adjacent parallel planes along the $[1\bar{2}10]$ direction.
In this configuration, the twin boundary $(10\bar{1}2)$ is edge on. 
The twin boundary migration is supposed to operate by the motion of $\vec{b}_{twin}=\vec{b}_{2/2}$ disconnection \cite{Cottrell1951, Westlake1961, Serra1999}. 
The notation $\vec{b}_{p/q}$ indicates that the disconnection has a step which height $h$ in the white and black lattice is $p$ and $q$ in units of $(10\bar{1}2)$ lattice planes spacing \cite{Braisaz1997}.
Following the notation of Frank and Pond their Burgers vector is \cite{Frank1965, Pond1987}:
\begin{equation}
\vec{b}_{twin}=\frac{1}{\lambda^2+2}\left( \begin{array}{c}
2+\lambda^2 \\ 
0 \\
-2-\lambda^2\\
\lambda(\lambda^2-2)
\end{array} \right)
\end{equation}

with $\lambda =\sqrt{\frac{2}{3}} \frac{c}{a}$ ($c$ and $a$ the lattice parameters).
The norm of the Burgers vector is $\| \vec{b}_{twin}\|=\frac{\sqrt{6}(2-\lambda^2)}{2\sqrt{\lambda^2+2}}a \approx 0.053$ nm,
and the step height $h=\frac{\sqrt{6}}{\sqrt{2+(2/\lambda)^2}}a \approx 0.265$ nm.
The coupling factor arising from the motion of this disconnection is then $\beta=\| \vec{b}_{twin} \| / h=\frac{2-\lambda^2}{\sqrt{2}\lambda}=0.1992$, in agreement with the measured value.

\section{Dislocation nucleation}
\begin{figure*}
\centering
\includegraphics[width=0.65\textwidth]{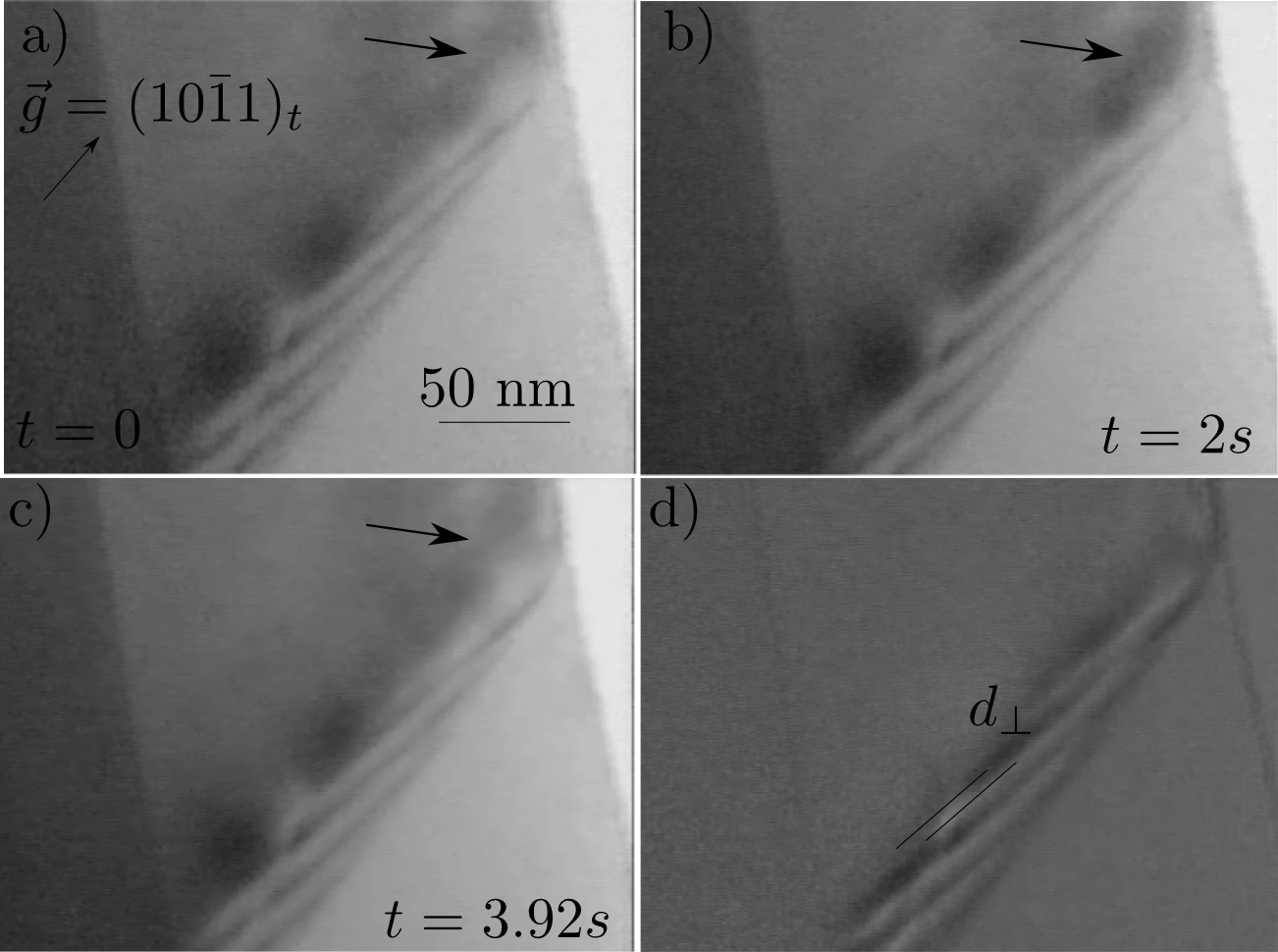}
\caption{Diffraction contrast broadening (indicated by the arrow) increased  from a) to b) and decreased from b) to c) near the fiber surface concomitantly with twin boundary migration. This can be interpreted by the emission of dislocations from the surface. Note the presence of secondary dislocations in the twin boundary dragged as the twin boundary migrated. d) is the image difference c)-a).}
\label{fig-twin-disloc}
\end{figure*}
A closer inspection of the twin boundary at higher magnification, under $\vec{g}=(10\bar{1}1)_t$, shows the presence of equally distributed dislocations with a spacing of 12 nm (figure \ref{fig-twin-disloc}). 
These dislocations are supposed to be secondary intrinsic dislocations that accommodate a slight misorientation from the $84.78^\circ$ perfect twin misorientation \cite{Bonnet1981}. 
Indeed, the observed twin boundary presents a misorientation $\approx 85.2^\circ$ around $[\bar{2}110]$, so that the extra misorientation can be accommodated by a single set of edge dislocations as observed.
During migration, these dislocations follow the twin boundary without moving laterally.
However, blurred and rapidly oscillating contrasts (see the video in the supplementary materials) can be clearly identified close to the fiber surface (indicated in figure \ref{fig-twin-disloc} by arrows).
The apparition of a strong contrast (fig. \ref{fig-twin-disloc}b) and its spreading (fig. \ref{fig-twin-disloc}c) along the twin boundary plane can be followed during few seconds.
Although difficult to interpret, it is associated with the motion of the twin boundary and appear only under stress, which is a strong indication of the nucleation of dislocation in an area close to the fiber surface. 
Figure \ref{fig-twin-disloc}d is the image difference c)-a). 
It shows that within the time interval where this strong contrast was observed, the twin boundary has progressed over a distance $d_\perp$ which can be determined by measuring the migration distance of the secondary dislocations.
The value of $d_\perp \approx 5.2 \pm 0.8$ nm approximately corresponds to the passage of 20 twinning dislocations, i.e. 20 times the step height $h=2d_{\{10\bar{1}2\}}=0.265$ nm.
The nucleation rate is $\dot{n} \approx 5$ dislocations per second, a value of the order of the dislocation production rate by spiral sources in Al fibers deformed homogeneously at low stress and strain rate ($\dot{\epsilon} \approx 3 \times 10^{-6}$ $s^{-1}$, $\tau \approx 100-200$ MPa) \cite{Mompiou2012b}.

\subsection{Twin nucleation}
\begin{figure*}
\centering
	\includegraphics[width=0.9\textwidth]{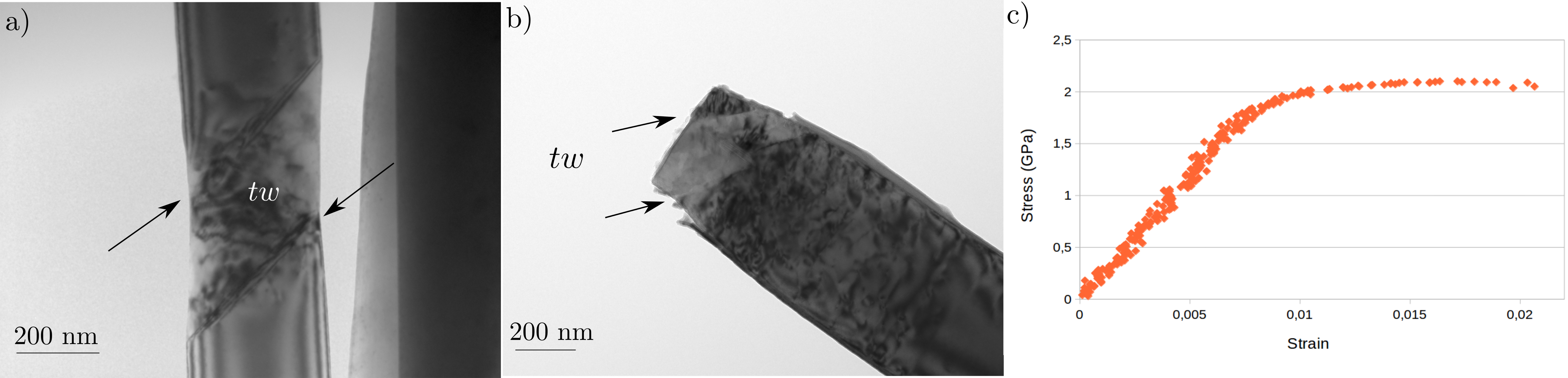}
	\caption{a) and b) are post-mortem evidence of twins ($tw$) that have formed inside a fiber. c) is the engineering stress strain curve corresponding to a micromechanical test performed on the fiber at a strain rate of $3 \times 10^{-5}$ $s^{-1}$ shown in b).}
\label{fig-twin-nucleation}
\end{figure*}

Although the first twin nucleation event has not been directly observed during in-situ experiments, figure \ref{fig-twin-nucleation} shows evidence of early twin formation and thickening.
Figure \ref{fig-twin-nucleation}a shows a twin lamella ($tw$) located somewhere along a fiber.
From both sides of the twin boundaries, indicated by arrows, no dislocations or clear defects can be identified.  Interestingly, a high dislocation density is visible in the twinned area, as expected from the orientation change. 
Indeed, whereas Schmid factors of prismatic slip systems has extremely low values  in the matrix oriented along the $\langle c \rangle$ axis, they reached almost their maximum in the twinned area.  
The plastic zone is thus mainly confined in the twinned area.
This observation tends to indicate that the twin has nucleated in the fiber oriented along its $\langle c \rangle$ axis, presumably from the surface, and eventually thickened through the entire fiber. 
This is reminiscent from the observation and MD simulation of long twin formation in gold nanowhiskers by the coalescence of individual twins nucleated from the surface \cite{Sedlmayr2012}.
Eventually plastic deformation due to prismatic slip has occurred.

Figure \ref{fig-twin-nucleation}b shows a twin ($tw$) near a fractured fiber.
Contrary to the case shown in figure \ref{fig-twin-nucleation}a,  the fiber is more heavily deformed, with twin boundaries distorted and numerous dislocations in the matrix.
This situation probably results a deformation at higher strain rate during which only a thin twinned area underwent ductile fracture rapidly.
In the absence of prismatic slip, as observed in figure \ref{fig-twin-migration}, the value of $\dot{\epsilon}_p \approx 5 \times 10^{-6}s^{-1}$, extracted from the twin boundary speed gives an order of magnitude of the strain rate for homogeneous deformation.

\section{Stress relaxation}
As twin boundary migration seems to operate by nucleation from the surface and propagation of twinning dislocations, we propose in this section a method to measure the corresponding activation volume.

\subsection{Activation volume measurement}

Macroscopically, the activation volume can be measured during a relaxation stress by plotting the variation of the strain rate logarithm versus the stress logarithm according to \cite{Caillard2003, Dotsenko1979}:
\begin{equation}
\label{eq-volume1}
V=kT \left. \frac{\partial \ln{\dot{\epsilon}_p}}{\partial \sigma} \right|_T
\end{equation}
If we consider that the dislocation density remain constant, the variation of the strain rate is equal to the variation of the dislocation speed. 
During in-situ experiment, the variation of the dislocation speed with the local stress, measured through the dislocation curvature, can be tentatively extracted leading to the determination of the activation volume.
In the present case, it is not possible to measure locally the stress but a clear decrease of the twin boundary speed during the relaxation is recorded. 
In the following, we discuss the possibility of extracting the activation volume from the measure of the twin boundary speed during the relaxation.

\subsection{Fundamental of stress relaxation}
\label{sect-fundamental}
From the above observations it can be supposed that a few fibers located in the right border of the hole  and which are not in the tangles are able to deform plastically. 
A careful examination of the machine stiffness given in Annex \ref{annex} indicates that the testing machine, composed of the copper grid, the glue, the straining rod and the thicker parts of the eutectic alloy, can be considered as almost infinitely hard compared to the specimen composed of fibers of total length $l_0 \approx 230 \pm 30$ $\mu$m. 
Under these conditions, the relation between the engineering strain, $\dot{\epsilon}$, engineering plastic strain, $\dot{\epsilon}_p$ and the observed true stress rate, $\dot{\sigma}$ is given in a good approximation \footnote{we consider that the plastic strain of the specimen is small and then the true plastic strain is equal to engineering plastic strain} by:
\begin{equation}
\dot{\epsilon} \approx \dot{\epsilon}_p+\frac{\dot{\sigma}}{E}
\label{eq1}
\end{equation}

For a relaxation test, the imposed displacement is null, i.e.\ $\dot{\epsilon}=0$ and then, 
\begin{equation}
\dot{\epsilon}_p\approx -\frac{\dot{\sigma}}{E}
\label{eq-relax}
\end{equation}

In the following we will assume that the relaxation curve i. e. $\sigma$ vs. $t$, exhibits a logarithmic decay of the stress with time. This has been proven to be true for a wide range of materials \cite{Caillard2003}:
\begin{equation}
\Delta \sigma=-\frac{kT}{V}\ln \left( 1+\frac{t-t_i}{t_0} \right)
\label{eq-log}
\end{equation}
with $V$ is the apparent activation volume and $t_0$ a time constant and $t_i$ the starting time of the relaxation.
Combining equations \ref{eq-relax} and \ref{eq-log} leads to:
\begin{equation}
\dot{\epsilon}_p \approx \frac{kT}{E V } \left(  \frac{1}{t+t_0-t_i} \right)
\label{eq-caillard-martin}
\end{equation}

\begin{figure}
\centering
	\includegraphics[width=0.2\textwidth]{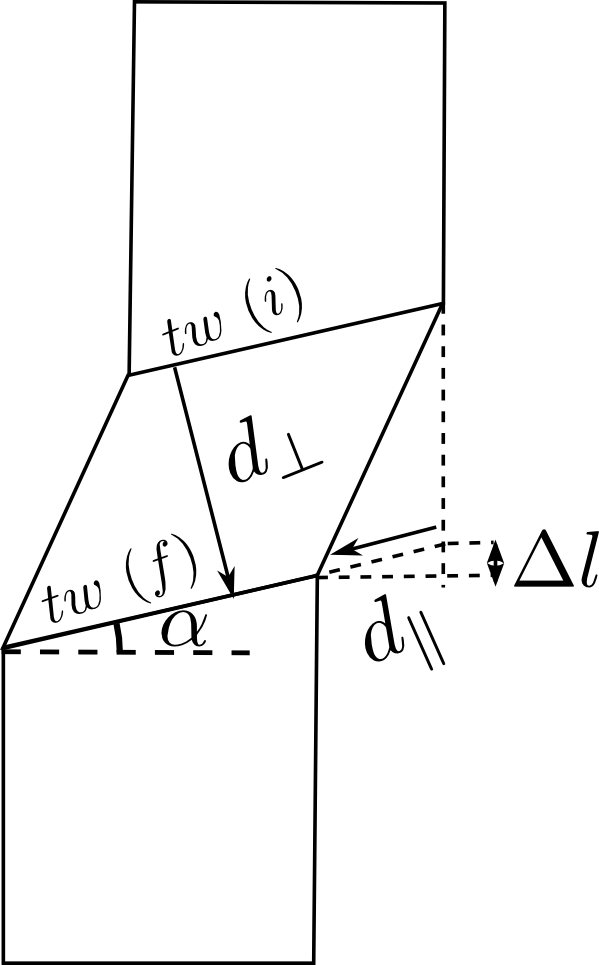}
	\caption{Schematics of the deformation of the fiber containing a twin boundary ($tw$) that migrates perpendicular to the twinning plane, inclined of an angle $90-\alpha$ with respect to the straining axis, over a distance $d_{\perp}$. This migration produces a shear displacement parallel to the twinning plane $d_{\|}$.}
\label{fig2}
\end{figure}

Now we have to establish the relation between the plastic strain rate and the twin boundary speed $v$.
Consider now the figure \ref{fig2} which schematically describes the evolution of the fiber under the migration of the twin boundary from its initial position ($tw \ (i)$) to its final one ($tw \ (f)$). During this motion over a distance $d_{\perp}$, a shear strain $d_{\|}$ is produced which leads to the elongation of the fiber.  

During a time interval $\Delta t$, the twin boundary migrates at a speed $v=\frac{d_{\perp}}{\Delta t}$, producing a shear strain $\beta=\frac{d_{\|}}{d_{\perp}}$. 
The specimen elongation is $\Delta l=d_{\|} \sin{\alpha}$, $\alpha$ being the twin boundary plane inclination and $l_0$ the gage length. 
The plastic strain is then:
\begin{equation}
\epsilon_p=\frac{\Delta l}{l_0}=\frac{d_{\|}\sin{\alpha}}{l_0}=\frac{\beta d_{\perp}\sin{\alpha}}{l_0}
\label{eq-eps}
\end{equation}
From equation \ref{eq-eps}, the plastic strain rate is given by:
\begin{equation}
\dot{\epsilon}_p=\frac{\beta d_{\perp} \sin{\alpha}}{l_0 \Delta t}=\frac{\beta \sin{\alpha}}{l_0}v
\label{eq-plast-strain-rate}
\end{equation}
The relation between the twin boundary speed and time can thus be derived combining equations \ref{eq-caillard-martin} and \ref{eq-plast-strain-rate}:
\begin{equation}
v=\frac{kT}{E V} \frac{l_0}{\beta \sin{\alpha}} \left( \frac{1}{t+t_0-t_i} \right)
\end{equation}
Then the activation volume can be derived by measuring the variation of the twin boundary speed with time.
	
\subsection{Stress relaxation experiment in a TEM}
\label{sect-exp}
\begin{figure}
\centering
	\includegraphics[width=0.45\textwidth]{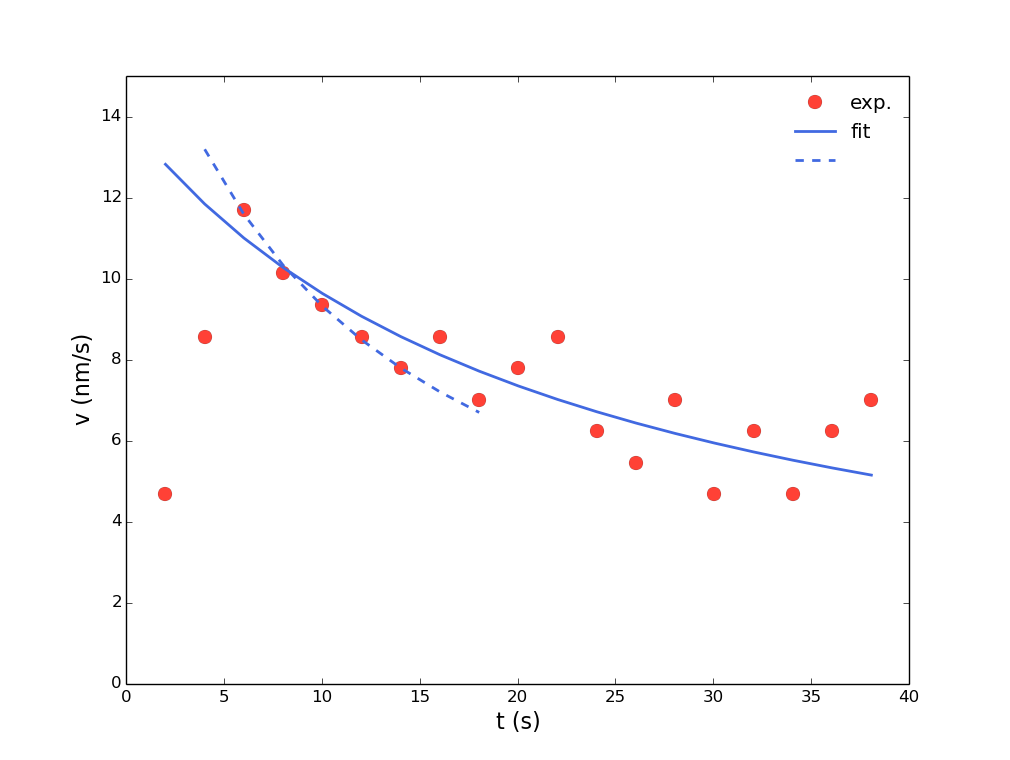}
	\caption{Variation of the twin boundary migration speed ($v$) as a function of  time ($t$). Lines indicate two data fit for data ranging between $t_0$ and $t$=40 s (full line) or $t$=15 s (dashed line) }
\label{fig3}
\end{figure}

During the stress relaxation, we have observed the motion of a twin boundary along the fiber, the speed of the twin boundary decreasing over the time.
The speed was measured using a custom Python script tracking the boundary trajectory \cite{pytrack}.
Figure \ref{fig3} shows the plot of the twin boundary speed vs. time. At $t\approx$ 0 s a displacement was imposed until the twin boundary moves. 
After the twin boundary starts to migrate ($t_i \approx 6$ s) the displacement was stopped.  The twin boundary continues to migrate with a speed decreasing which can be fitted by the equation: 
\begin{equation}
v=\frac{a_1}{t+t_0-t_i}
\label{eq-fit}
\end{equation}
The two lines in figure \ref{fig3} are possible data fit when selecting different time intervals, from $t_i$ to 40 s in for the full line and from $t_i$ to 15 s for the dashed line (early stage of the relaxation).
In the overall, the fitting parameters are $t_0$= 10 $\pm$ 6 s, $a_1=$250 $\pm$ 60 nm.
Combining equations \ref{eq-fit} and \ref{eq-plast-strain-rate} gives an estimate of the activation volume:
\begin{equation}
V=\frac{kTl_0}{Ea_1\beta \sin{\alpha}}
\end{equation}
Taking $T=300$ K ($kT=4\times 10^{-21}$ J), $\beta=0.2$, $E=287$ GPa, $\alpha=53^\circ$, $l_0 \approx 230 \pm 30$ $\mu$m leads to $V \approx 8.3 \pm 3.3 \times 10^{-29}$ m$^3$, which represents $V \approx 574 \pm 226 b^3$.
This value is few times larger than the ones found for nanotwinned copper $V_{nano-twin}=12-20b^3$ \cite{Lu2009b} and nanocrystalline aluminium  \cite{Gianola2006a} $V_{ncAl}=10-55 b^3$, where plastic deformation involves mainly twin or grain boundaries.

\subsection{Activation energy evaluation}
Although the activation energy $G$ cannot be measured directly here, an estimation can be made. 
Considering that the nucleation of twinning dislocations is thermally activated, the nucleation time $t$ at the temperature $T$ is given by:
\begin{equation}
\frac{1}{t}= A \exp \left( -\frac{G}{kT} \right)
\end{equation}
where $A$ is a constant related to the nucleation attempt frequency and to the entropy \cite{Hirel2008}.
Considering that $1 \times 10^{11} s^{-1}<A \approx \nu^{\star}<5 \times 10^{11} s^{-1}$ (cf. \ref{annex2}), and that $1/t=v/h$ yields at $T=300$ K 0.56 eV$<G<$0.72 eV, i.e. around 25 kT.

\section{Discussion}
\label{sect-discuss}
The activation volume estimated above can be tentatively explained by considering the different mechanisms involved in the twin nucleation and propagation.
The activation volume is related to the variation of the free energy $G$ associated with the process.
\begin{equation}
V=-\frac{\partial G}{\partial \sigma}
\label{eq-volume2}
\end{equation}

\subsection{Twinning dislocation nucleation}
\begin{figure*}
\centering
	\includegraphics[width=\textwidth]{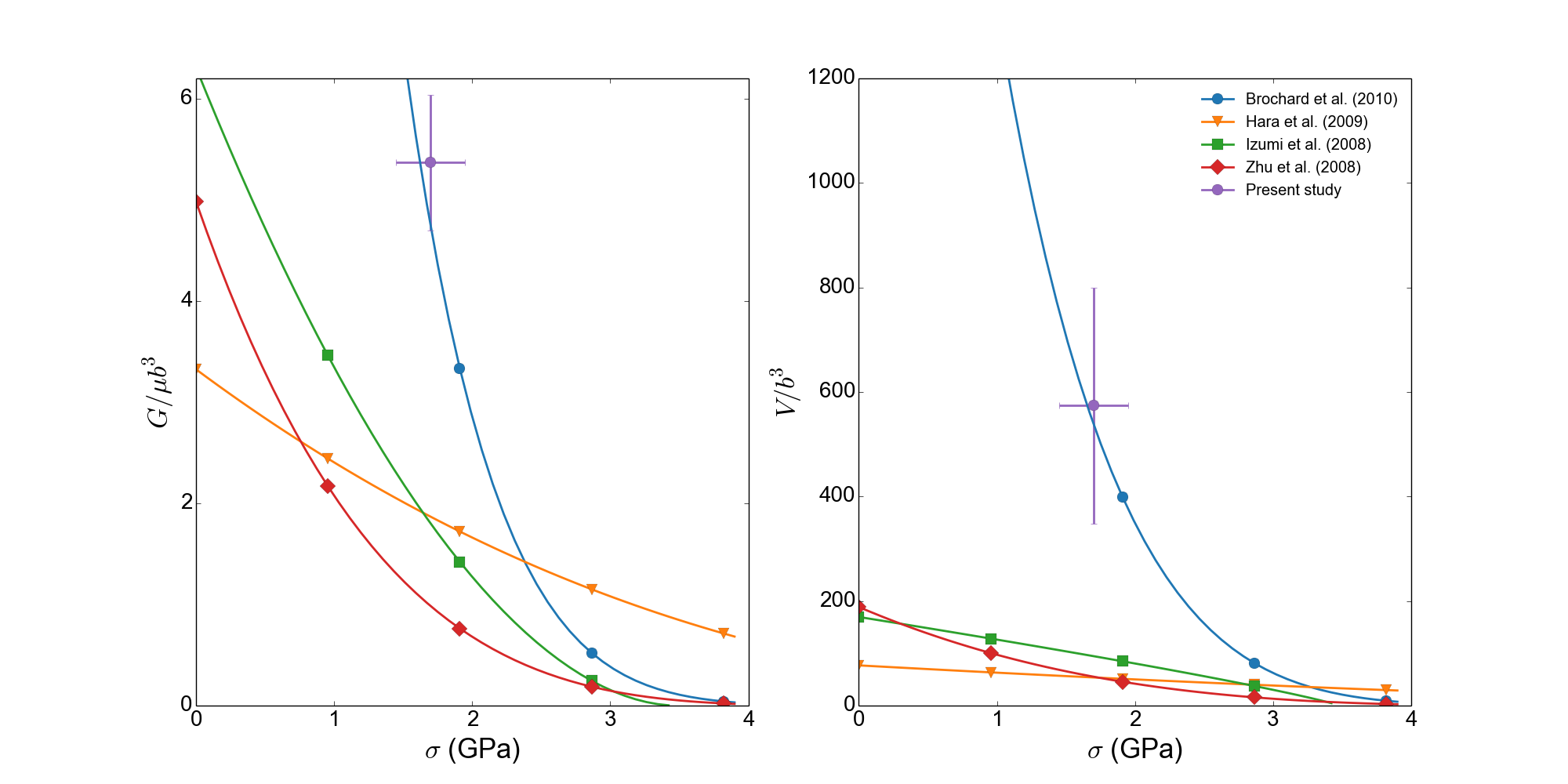}
	\caption{Normalized activation energy (a) and activation volume (b) as a function of the stress. Full lines correspond to results of atomistic simulations of dislocation surface nucleation. The values of the present study are also indicated for comparison.}
\label{fig-volume}
\end{figure*}
The nucleation of a dislocation embryo from either  a flat  \cite{Hara2009} or stepped surface \cite{Brochard2010,Hirel2008, Hara2009}, or from a corner \cite{Izumi2008, Zhu2008} have been extensively studied by atomistic simulations and elastic calculations \cite{Hirel2008}.
All these studies conclude that the activation energy takes the form:
\begin{equation}
\label{eq-E}
G(T, \sigma)=G_0 \left(1-T/T_m \right) \left( 1-\sigma/\sigma_{ath}\right)^\alpha
\end{equation}
where $G_0$ is a constant, $T_m $ is the surface disordering temperature \cite{Zhu2008} taken as half the melting temperature, $\sigma_{ath}$ the athermal stress and $\alpha$ a constant.
The variation of the activation energy, normalized by $\mu b^3$, and the activation volume, normalized by $b^3$, as a function of stress have been redrawn from these studies as dotted lines in figure \ref{fig-volume}a and b, respectively.
Although the stress was not retrieved in the present study, the yield stress of the fiber shown in figure \ref{fig-twin-nucleation}b has been estimated by micromechanical testing to $\sigma \approx 1.7 \pm 0.25$ GPa (see the stress-strain curve in fig. \ref{fig-twin-nucleation}c) for a strain rate of $\dot{\epsilon} \approx 7 \times 10^{-3}$ s$^{-1}$.
It can be shown in \ref{annex2} that for surface nucleation $\sigma(\dot{\epsilon}_p=5 \times 10^{-6}) \approx \sigma(\dot{\epsilon}_p=7 \times 10^{-3})$ at $T=300$ K.
The value of activation energy and volume measured in the present study is reported in figure  \ref{fig-volume}.
Despite the scatter in the activation energy and volume from the atomistic studies due to the importance of the geometrical factor in the nucleation process \cite{Brochard2010}, the experimental values $5 \times 10^{-29}$ m$^3$ ($348b^3$)$<V<1.2 \times 10^{-28}$ m$^3$ (800$b^3$),  although slightly higher compared to the values obtained by most of the atomistic simulations, are still comparable for instance with the work of \cite{Brochard2010}.
It is interesting to note here that if the migration of the twin boundary is due to surface nucleation, it should not be influenced by the presence of specific nucleation site, such as a step, because of the continuous and regular migration of the twin boundary.

\subsection{Atomic shuffling during twinning dislocation motion}

While moving laterally, atoms below the disconnection step have to be shuffled. 
This mechanism is supposed to be thermally activated so that the activation area associated corresponds to the area where atoms have to be rearranged (in red in figure \ref{fig-dichro}). This area is $A_r=h \times d_t$ where $h=2d_{(10\bar{1}2)}=0.265$ nm is the step height. 
The activation volume is then given by $V_{shuffling}=A_r \times \| \vec{b}_{twin}\|=7.5 \times 10^{-30}$ m$^3$ which is significantly smaller than the volume measured.

\subsection{Twinning dislocation interaction with solute atoms}
Because the fiber may not be pure, solute atoms can act as pinning points on the twinning dislocation. 
In that case, the activation volume is given by \cite{Caillard2003}:
\begin{equation}
V_{solute}=b^3/c_{solute}
\end{equation}
where $c_{solute}$ is the atomic concentration of solute atoms.
According to the Al-Be phase diagram \cite{Okamoto2000}, the concentration of Al solute atoms is expected to be much less than 60 ppm which leads to an activation volume $V_{solute}>2 \times 10^{-26}$ m$^{-3}$ much larger than expected.
Although O atoms can also be present, their concentration is also presumably very low, ruling out the interaction of twinning dislocations with solute atoms as the controlling mechanism.

\subsection{Friction stress acting on a twinning dislocation}

In most hcp metals, dislocation motion is controlled by a friction force (Peierls-Nabarro stress) due to  non planar dislocation core dissociation \cite{Caillard2003}. 
In beryllium, this friction stress leads to the locking and unlocking of screw dislocations in the prismatic plane. The stress anomaly in the range $80 -450$ K has been explained by this phenomenon \cite{Couret1989}.
Although, this topic is poorly documented, twinning dislocation can also be sensible to a friction stress along the twin plane. 
Serra et al. have indeed shown that the size of twinning dislocation core in hcp metals greatly  influences their mobility \cite{Serra1991}.
Similar conclusions have been drawn for rhombohedral twinning in sapphire where measured activation volume  and enthalpy fit with a model of double kink pair nucleation along the twinning dislocation \cite{Choi1995}.
Serra et al. \cite{Serra1991} have shown however that for $\{10\bar{1}2\}$ twin, twinning dislocation can move at 0 K at a small strain of 0.002, which gives in our case, considering a Schmid factor of $s_f=0.5$, a friction stress $\sigma= 0.002 E/s_f  \approx 0.5$ GPa, smaller than the yield stress. This indicates that the motion of a twinning dislocation through the nucleation and propagation of kinks along its line \cite{Hirth1982,Caillard2003} is athermal in our observation and thus associated with activation volume equal to zero.

\section{Conclusions}
Tensile in-situ TEM experiments were performed on sub-micron beryllium fibers containing a low initial dislocation content. They revealed the thickening of a twin domain that occurred through the migration of a $\{10\bar{1}2\}$ twin boundary under stress. 
The motion of this twin boundary appeared continuous and associated with rapidly oscillating contrasts near the surface.
The absence of dislocation in the fiber interior discard a twinning mechanism operating through a dislocation source (pole mechanism).
The shear strain associated have been measured and found in agreement with the value expected from the motion of zonal twinning dislocations. We have therefore attributed this rapid contrast oscillation to repeated dislocation nucleation.
In parallel, we provide a method to retrieve the activation volume associated with the twin boundary migration by measuring the twin boundary speed decrease during the stress relaxation. 
We found an experimental value of  $8.3 \pm 3.3 \times 10^{-29}$ m$^3$, which  is comparable to the value expected from dislocation surface nucleation.
These experiments thus provide the first combination of a dynamic observation of a single migrating twin boundary in a sub-micron Be wire and the quantitative measurements of activation parameters associated with this movement. We show that the migration depends exclusively on the surface nucleation of dislocation at the location of the boundary.
Further experiments focusing on the measure of activation parameters as a function of stress through micromechanical testing of individual fibers may provide a complete picture of this mode of deformation that is probably peculiar to sub-micron size crystals. Along with the present results, they should provide an excellent bench test for dedicated atomistic simulations and gain direct insights into the controlling mechanism ruling the plastic deformation of micro- and nanometer scale crystals.

\section*{Acknowledgement}
The authors would like to thank D. Caillard and S. Brochard for valuable discussions. They acknowledge J. Bonneville and A. Serra for their comments after the initial publication. M.L. would also like to thank the Humbolt Fundation for allowing him to spend time at KIT. C.E. and O.K. acknowledge the support by the Robert-Bosch Foundation. This work has been supported by the French National Research Agency under the "Investissement d'Avenir" program reference No. ANR-10-EQPX-38-01.
\appendix
\section{Effective modulus of the sample-machine system}
\label{annex}
The relation between the engineering strain, $\dot{\epsilon}$, engineering plastic strain, $\dot{\epsilon}_p$ and the observed true stress rate, $\dot{\sigma}$ is given by:
\begin{equation}
\dot{\epsilon} \approx \dot{\epsilon_p}+\frac{\dot{\sigma}}{C}
\label{eq0}
\end{equation}
In this equation, C is the effective modulus of the sample-machine system:
\begin{equation}
\frac{1}{C}=\frac{1}{E}+\frac{1}{M}
\label{eq2}
\end{equation}
where $E$ is the Young's modulus of the specimen and $M$ is 
\begin{equation}
M=K\frac{l_0}{A_0}
\label{eq3}
\end{equation}
and $K$ the machine stiffness.
The machine stiffness is a combination of the different stiffness of the elements composing the machine (see figure \ref{fig-gage}). 

For elements deforming in tension:
\begin{equation}
\frac{1}{K}=\sum_{i} \frac{1}{K_i}=\sum_{i} \frac{l_i}{E_i A_i}
\label{eq4}
\end{equation}
with $E_i$, $A_i$ and $l_i$ are the Young's modulus, the cross-section area and the length of the $i^{th}$ element, respectively.
Because the sample is glued on the copper grid, the cyanoacrylate glue also deforms mainly by shear. The shear stiffness of the glue is:
\begin{equation}
\frac{1}{K_{glue}} =\frac{t_{glue}}{G_{glue}A_{glue}}
\label{eq-stiff-glue}
\end{equation}
With $t_{glue}$ the thickness of the glue film, $G_{glue}$ the shear modulus and $A_{glue}$ the surface sheared.
Combining equations \ref{eq2},  \ref{eq4} and \ref{eq-stiff-glue} 	leads to:
\begin{equation}
\frac{1}{C}=\frac{1}{E}+\frac{A_0}{l_0}\left(  \sum_{i} \frac{l_i}{E_i A_i} +\frac{t_{glue}}{G_{glue}A_{glue}}\right)
\label{eq-stiff}
\end{equation}

In the present situation, equation \ref{eq-stiff} is the combination of the stiffness of the tensile rod, of the copper grid, of the eutectic alloy, of the glue (see figure \ref{fig-gage}):
\begin{equation}
\begin{split}
\label{eq10}
\frac{1}{C}&=\frac{1}{E}+\frac{A_0}{l_0} \left( \frac{l_{rod}}{E_{rod} A_{rod}}+\frac{l_{grid}}{E_{grid} A_{grid}} \right. \\
&\quad \left. +\frac{l_{eutectic}}{E_{eutectic} A_{eutectic}}+ \frac{t_{glue}}{G_{glue}A_{glue}} \right)
\end{split}
\end{equation}
The fiber gage length measured from TEM observation is $l_0=230 \pm 30 \  \mu$m. 
Assuming the fiber cross-section is a disk of diameter $d_f$=400 nm, the cross section area $A_f \approx \pi d_f^2/4$. Considering that 4 fibers deform plastically, the specimen cross-section is $A_0 \approx 4A_f \approx \pi d_f^2$. 
The straining rod is cylindrical with a cross sectional area $A_{rod}=\pi d_{rod}^2/4$, with $d_{rod}=3$ mm. 
The rod length is about $30$ cm. 
The Young's modulus of the rod is close to $E_{rod}\approx 200 GPa =0.7 E$. 
The copper grid cross section area is $A_{grid}=t_{grid}w_{grid}$ with $t_{grid}=70 \ \mu$m the grid thickness and $w_{grid}=2.3$ mm the grid width. 
The grid length is about $8$ mm.
The Young's modulus of the grid is $E_{grid}=117$ GPa$=0.4 E$.
The part of the eutectic which has not been etched is also subjected to deformation. 
The  stress will concentrate in the middle of the hole rim and in area perpendicular to the straining axis as shown in \cite{Coujou1990}. The hole rim is however not thin as a typical TEM specimen, so that the strain should be localized in the fibers. The thickness of the eutectic part is about 50 $\mu$m, 1 mm large and about 3 mm long.
The Young's modulus of the eutectic can be estimated using the rule of mixture $E_c=\sum_i V_i E_i$ with $V_i=\frac{W_i/\rho_i}{\sum_i W_i/\rho_i} $ the volume fraction of the i$^{th}$ element of weight fraction $W_i$, density $\rho_i$ and Young's modulus $E_i$. Since there is approximately 1.5 wt.$\%$ Be in the eutectic, the effective modulus is $E_{eutectic} \approx 0.98E_{Al}+0.02E_{Be}\approx 74 GPa \approx 0.26E$.
The thickness of the glue is estimated to $t_g \approx 10\ \mu$m. The area covered by the glue is approximately 1 mm large (the sample width) by 0.5 mm. The shear modulus of the glue is of the of 1 GPa i.e. $G_g \approx 0.005E$.
We finally obtain from eq. \ref{eq10}:
\begin{equation}
\begin{split}
\frac{1}{C} & \approx \frac{1}{E}+\frac{1.5\times10^{-4}}{E}+\frac{3.2\times10^{-4}}{E} +\frac{3.2\times10^{-4}}{E} \\
& +\frac{8\times10^{-6}}{E} \approx \frac{1}{E} 
\end{split}
\end{equation}

Then, equation \ref{eq0} can be replaced by equation \ref{eq1}.

\section{Variation of the stress with strain rate for surface nucleation}
\label{annex2}
The variation of the stress with the plastic strain rate can be estimated in the case of surface nucleation by combining Eq. \ref{eq-volume1}, \ref{eq-volume2} and \ref{eq-E}, and by integration with respect to the stress between $\sigma$ and $\sigma_{ath}$:
\begin{equation}
\label{eq-sigma}
\sigma=\sigma_{ath} \left( 1- \left( \frac{kT}{E_0(1-T/T_m)} \ln \frac{C \nu^\star}{\dot{\epsilon}_p}\right) ^{1/\alpha} \right)
\end{equation}
where $\nu^\star$ is a characteristic nucleation frequency and $C=n_s \cos(\theta)$ a geometrical factor that relates the nucleation rate to the plastic strain rate ($n_s$ being the number of dislocation surface sources per fiber unit length and $\theta$ the angle between the glide plane and the fiber surface).
Taking the values given in \cite{Brochard2010} for Al, i.e. $T=300$ K, $T_m=800$ K, $E_0=499.6$ eV, $\nu^\star=10^{11}$ $s^{-1}$, $\alpha=7.02$, and $C=0.011$ leads to $\sigma(7 \ 10^{-3})/\sigma(5 \ 10^{-6}) =1.03$.
For Cu and according to \cite{Zhu2008}, taking $T=300$ K, $T_m=692$ K, $E_0=4.8$ eV, $\nu^\star=5.02 \ 10^{13}$ $s^{-1}$, $\alpha=4.1$ and $C=0.012$, leads to $\sigma(7 \ 10^{-3})/\sigma(5 \ 10^{-6}) =1.17$.
In Be, although the activation parameters have not been determined, we can make the following assumptions: $\nu^\star=\nu_D=3 \times 10^{-13}$ $s^{-1}$, the Debye frequency, $T_m=800$ K, $C=0.012$, $\alpha$ ranging from 4 to 7 and $E_0$ ranging from 4 to 500 eV. In all cases, this yields also to $1<\sigma(7 \ 10^{-3})/\sigma(5 \ 10^{-6}) <1.2$.



\bibliography{/home/fred/Documents/papiers/ref.bib} 

\begin{thebibliography}{71}
\expandafter\ifx\csname natexlab\endcsname\relax\def\natexlab#1{#1}\fi
\providecommand{\bibinfo}[2]{#2}
\ifx\xfnm\relax \def\xfnm[#1]{\unskip,\space#1}\fi
\bibitem[{Sharon et~al.(2014)Sharon, Zhang, Mompiou, Legros, and
  Hemker}]{Sharon2014}
\bibinfo{author}{J.~Sharon}, \bibinfo{author}{Y.~Zhang},
  \bibinfo{author}{F.~Mompiou}, \bibinfo{author}{M.~Legros},
  \bibinfo{author}{K.~Hemker}, \bibinfo{journal}{Scripta Materialia}
  \bibinfo{volume}{75} (\bibinfo{year}{2014}) \bibinfo{pages}{10 -- 13}.
\bibitem[{Christian and Mahajan(1995)}]{Christian1995}
\bibinfo{author}{J.~Christian}, \bibinfo{author}{S.~Mahajan},
  \bibinfo{journal}{Progress in Materials Science} \bibinfo{volume}{39}
  (\bibinfo{year}{1995}) \bibinfo{pages}{1--157}.
\bibitem[{Yoo(1981)}]{Yoo1981}
\bibinfo{author}{M.~Yoo}, \bibinfo{journal}{Metallurgical Transactions A}
  \bibinfo{volume}{12} (\bibinfo{year}{1981}) \bibinfo{pages}{409--418}.
\bibitem[{Gutierrez-Urrutia and Raabe(2012)}]{Gutierrez2012}
\bibinfo{author}{I.~Gutierrez-Urrutia}, \bibinfo{author}{D.~Raabe},
  \bibinfo{journal}{Acta Materialia} \bibinfo{volume}{60}
  (\bibinfo{year}{2012}) \bibinfo{pages}{5791--5802}.
\bibitem[{Kalidindi et~al.(2003)Kalidindi, Salem, and Doherty}]{Kalidindi2003}
\bibinfo{author}{S.~R. Kalidindi}, \bibinfo{author}{A.~A. Salem},
  \bibinfo{author}{R.~D. Doherty}, \bibinfo{journal}{Advanced Engineering
  Materials} \bibinfo{volume}{5} (\bibinfo{year}{2003})
  \bibinfo{pages}{229--232}.
\bibitem[{Lu et~al.(2009)Lu, Chen, Huang, and Lu}]{Lu2009}
\bibinfo{author}{L.~Lu}, \bibinfo{author}{X.~Chen}, \bibinfo{author}{X.~Huang},
  \bibinfo{author}{K.~Lu}, \bibinfo{journal}{Science} \bibinfo{volume}{323}
  (\bibinfo{year}{2009}) \bibinfo{pages}{607--610}.
\bibitem[{Bouaziz et~al.(2011)Bouaziz, Allain, Scott, Cugy, and
  Barbier}]{Bouaziz2011}
\bibinfo{author}{O.~Bouaziz}, \bibinfo{author}{S.~Allain},
  \bibinfo{author}{C.~Scott}, \bibinfo{author}{P.~Cugy},
  \bibinfo{author}{D.~Barbier}, \bibinfo{journal}{Curr. Opinion in Sol.State
  and Mat.Science} \bibinfo{volume}{15} (\bibinfo{year}{2011})
  \bibinfo{pages}{141 -- 168}.
\bibitem[{Zhu et~al.(2012)Zhu, Liao, and Wu}]{Zhu2012}
\bibinfo{author}{Y.~Zhu}, \bibinfo{author}{X.~Liao}, \bibinfo{author}{X.~Wu},
  \bibinfo{journal}{{Prog. Materials Science}} \bibinfo{volume}{{57}}
  (\bibinfo{year}{{2012}}) \bibinfo{pages}{{1--62}}.
\bibitem[{Liu et~al.(2014)Liu, Wang, Li, Lu, Zhang, Shan, Li, Jia, Sun, and
  Ma}]{Liu2014}
\bibinfo{author}{B.-Y. Liu}, \bibinfo{author}{J.~Wang},
  \bibinfo{author}{B.~Li}, \bibinfo{author}{L.~Lu}, \bibinfo{author}{X.-Y.
  Zhang}, \bibinfo{author}{Z.-W. Shan}, \bibinfo{author}{J.~Li},
  \bibinfo{author}{C.-L. Jia}, \bibinfo{author}{J.~Sun},
  \bibinfo{author}{E.~Ma}, \bibinfo{journal}{Nature communications}
  \bibinfo{volume}{5} (\bibinfo{year}{2014}).
\bibitem[{Chassagne et~al.(2011)Chassagne, Legros, and Rodney}]{Chassagne2011}
\bibinfo{author}{M.~Chassagne}, \bibinfo{author}{M.~Legros},
  \bibinfo{author}{D.~Rodney}, \bibinfo{journal}{Acta Materialia}
  \bibinfo{volume}{59} (\bibinfo{year}{2011}) \bibinfo{pages}{1456--1463}.
  \bibinfo{note}{Doi: DOI: 10.1016/j.actamat.2010.11.007}.
\bibitem[{Morrow et~al.(2014)Morrow, McCabe, Cerreta, and Tome}]{Morrow2014}
\bibinfo{author}{B.~Morrow}, \bibinfo{author}{R.~McCabe},
  \bibinfo{author}{E.~Cerreta}, \bibinfo{author}{C.~Tome},
  \bibinfo{journal}{Metallurgical and Materials Transactions A}
  \bibinfo{volume}{45} (\bibinfo{year}{2014}) \bibinfo{pages}{36--40}.
\bibitem[{Caillard and Martin(2003)}]{Caillard2003}
\bibinfo{author}{D.~Caillard}, \bibinfo{author}{J.~Martin},
  \bibinfo{title}{{Thermally activated mechanisms in crystal plasticity}},
  volume~\bibinfo{volume}{1}, \bibinfo{publisher}{Pergamon},
  \bibinfo{address}{Cambridge}, \bibinfo{year}{2003}.
\bibitem[{Lu et~al.(2009)Lu, Zhu, Shen, Dao, Lu, and Suresh}]{Lu2009b}
\bibinfo{author}{L.~Lu}, \bibinfo{author}{T.~Zhu}, \bibinfo{author}{Y.~Shen},
  \bibinfo{author}{M.~Dao}, \bibinfo{author}{K.~Lu},
  \bibinfo{author}{S.~Suresh}, \bibinfo{journal}{Acta Materialia}
  \bibinfo{volume}{57} (\bibinfo{year}{2009}) \bibinfo{pages}{5165--5173}.
\bibitem[{Mompiou et~al.(2012)Mompiou, Caillard, Legros, and
  Mughrabi}]{Mompiou2012}
\bibinfo{author}{F.~Mompiou}, \bibinfo{author}{D.~Caillard},
  \bibinfo{author}{M.~Legros}, \bibinfo{author}{H.~Mughrabi},
  \bibinfo{journal}{Acta Mater} \bibinfo{volume}{60} (\bibinfo{year}{2012})
  \bibinfo{pages}{3402 -- 3414}.
\bibitem[{Kwiecinski and Wyrzykowski(1991)}]{Kwiecinski1991a}
\bibinfo{author}{J.~Kwiecinski}, \bibinfo{author}{J.~W. Wyrzykowski},
  \bibinfo{journal}{Acta Metallurgica et Materialia} \bibinfo{volume}{39}
  (\bibinfo{year}{1991}) \bibinfo{pages}{1953}.
\bibitem[{Poirier et~al.(1967)Poirier, Antolin, and Dupouy}]{Poirier1967}
\bibinfo{author}{J.~Poirier}, \bibinfo{author}{J.~Antolin},
  \bibinfo{author}{J.~Dupouy}, \bibinfo{journal}{Can. J. Phys.}
  \bibinfo{volume}{45} (\bibinfo{year}{1967}) \bibinfo{pages}{1221-- 1223}.
\bibitem[{Regnier and Dupouy(1970)}]{Regnier1970}
\bibinfo{author}{P.~Regnier}, \bibinfo{author}{J.~M. Dupouy},
  \bibinfo{journal}{Phys. Stat. Sol.} \bibinfo{volume}{39}
  (\bibinfo{year}{1970}) \bibinfo{pages}{79--93}.
\bibitem[{Beuers et~al.(1987)Beuers, Jonsson, and Petzow}]{Beuers1987}
\bibinfo{author}{J.~Beuers}, \bibinfo{author}{S.~Jonsson},
  \bibinfo{author}{G.~Petzow}, \bibinfo{journal}{Acta Metall}
  \bibinfo{volume}{35} (\bibinfo{year}{1987}) \bibinfo{pages}{2277--2287}.
\bibitem[{Couret and Caillard(1989)}]{Couret1989}
\bibinfo{author}{A.~Couret}, \bibinfo{author}{D.~Caillard},
  \bibinfo{journal}{Phil. Mag. A} \bibinfo{volume}{59} (\bibinfo{year}{1989})
  \bibinfo{pages}{783--800}.
\bibitem[{Cottrell and Bilby(1951)}]{Cottrell1951}
\bibinfo{author}{A.~Cottrell}, \bibinfo{author}{B.~Bilby},
  \bibinfo{journal}{Phil. Mag.} \bibinfo{volume}{42} (\bibinfo{year}{1951})
  \bibinfo{pages}{573--581}.
\bibitem[{Thompson and Millard(1952)}]{Thompson1952}
\bibinfo{author}{N.~Thompson}, \bibinfo{author}{D.~Millard},
  \bibinfo{journal}{Phil. Mag.} \bibinfo{volume}{43} (\bibinfo{year}{1952})
  \bibinfo{pages}{422--440}.
\bibitem[{Westlake(1961)}]{Westlake1961}
\bibinfo{author}{D.~Westlake}, \bibinfo{journal}{Acta Metall.}
  \bibinfo{volume}{9} (\bibinfo{year}{1961}) \bibinfo{pages}{327--331}.
\bibitem[{Serra et~al.(1988)Serra, Bacon, and Pond}]{Serra1988}
\bibinfo{author}{A.~Serra}, \bibinfo{author}{D.~Bacon},
  \bibinfo{author}{R.~Pond}, \bibinfo{journal}{Acta Metallurgica}
  \bibinfo{volume}{36} (\bibinfo{year}{1988}) \bibinfo{pages}{3183 -- 3203}.
\bibitem[{Hirth and Pond(1996)}]{Hirth1996}
\bibinfo{author}{J.~P. Hirth}, \bibinfo{author}{R.~C. Pond},
  \bibinfo{journal}{{Acta Mater.}} \bibinfo{volume}{{44}}
  (\bibinfo{year}{{1996}}) \bibinfo{pages}{{4749--4763}}.
\bibitem[{Price(1960)}]{Price1960}
\bibinfo{author}{P.~B. Price}, \bibinfo{journal}{Proc Roy Soc A}
  \bibinfo{volume}{260} (\bibinfo{year}{1960}) \bibinfo{pages}{251--262}.
\bibitem[{Roos et~al.(2014)Roos, Kapelle, Richter, and Volkert}]{Roos2014}
\bibinfo{author}{B.~Roos}, \bibinfo{author}{B.~Kapelle},
  \bibinfo{author}{G.~Richter}, \bibinfo{author}{C.~A. Volkert},
  \bibinfo{journal}{Applied Physics Letters} \bibinfo{volume}{105}
  (\bibinfo{year}{2014}) \bibinfo{pages}{--}.
\bibitem[{Yu et~al.(2012)Yu, Chen, Mishra, and Li}]{Yu2012a}
\bibinfo{author}{L.~Yu, Q.and~Qi}, \bibinfo{author}{K.~Chen},
  \bibinfo{author}{R.~K. Mishra}, \bibinfo{author}{A.~M. Li, J.and~Minor},
  \bibinfo{journal}{Nano Letters} \bibinfo{volume}{12} (\bibinfo{year}{2012})
  \bibinfo{pages}{887--892}.
\bibitem[{Narayan and Zhu(2008)}]{Narayan2008}
\bibinfo{author}{J.~Narayan}, \bibinfo{author}{Y.~T. Zhu},
  \bibinfo{journal}{Applied Physics Letters} \bibinfo{volume}{92}
  (\bibinfo{year}{2008}) \bibinfo{pages}{--}.
\bibitem[{Shu et~al.(2013)Shu, Liu, Shen, and Hu}]{Shu2013}
\bibinfo{author}{B.~Shu}, \bibinfo{author}{L.~Liu}, \bibinfo{author}{B.~Shen},
  \bibinfo{author}{W.~Hu}, \bibinfo{journal}{Materials Letters}
  \bibinfo{volume}{106} (\bibinfo{year}{2013}) \bibinfo{pages}{225 -- 228}.
\bibitem[{Sennour et~al.(2007)Sennour, Lartigue-Korinek, Champion, and
  Hytch}]{Sennour2007}
\bibinfo{author}{M.~Sennour}, \bibinfo{author}{S.~Lartigue-Korinek},
  \bibinfo{author}{Y.~Champion}, \bibinfo{author}{M.~J. Hytch},
  \bibinfo{journal}{{Phil. Mag.}} \bibinfo{volume}{{87}}
  (\bibinfo{year}{{2007}}) \bibinfo{pages}{{1465--1486}}.
\bibitem[{Chen et~al.(2003)Chen, Ma, Hemker, Sheng, Wang, and Cheng}]{Chen2003}
\bibinfo{author}{M.~Chen}, \bibinfo{author}{E.~Ma},
  \bibinfo{author}{K.~Hemker}, \bibinfo{author}{H.~Sheng},
  \bibinfo{author}{Y.~Wang}, \bibinfo{author}{X.~Cheng},
  \bibinfo{journal}{Science} \bibinfo{volume}{300} (\bibinfo{year}{2003})
  \bibinfo{pages}{1275--7}.
\bibitem[{Van~Swygenhoven et~al.(2006)Van~Swygenhoven, Derlet, and
  Froseth}]{VanSwygenhoven2006}
\bibinfo{author}{H.~Van~Swygenhoven}, \bibinfo{author}{P.~Derlet},
  \bibinfo{author}{A.~Froseth}, \bibinfo{journal}{Acta Mater.}
  \bibinfo{volume}{54} (\bibinfo{year}{2006}) \bibinfo{pages}{1975--1983}.
\bibitem[{Godet et~al.(2004)Godet, Pizzagalli, Brochard, and
  Beauchamp}]{Godet2004}
\bibinfo{author}{J.~Godet}, \bibinfo{author}{L.~Pizzagalli},
  \bibinfo{author}{S.~Brochard}, \bibinfo{author}{P.~Beauchamp},
  \bibinfo{journal}{Phys. Rev. B} \bibinfo{volume}{70} (\bibinfo{year}{2004})
  \bibinfo{pages}{054109}.
\bibitem[{Guenole et~al.(2013)Guenole, Godet, and Brochard}]{Guenole2013}
\bibinfo{author}{J.~Guenole}, \bibinfo{author}{J.~Godet},
  \bibinfo{author}{S.~Brochard}, \bibinfo{journal}{Phys. Rev. B}
  \bibinfo{volume}{87} (\bibinfo{year}{2013}) \bibinfo{pages}{045201}.
\bibitem[{Yu et~al.(2010)Yu, Shan, Li, Huang, Xiao, Sun, and Ma}]{Yu2010}
\bibinfo{author}{Q.~Yu}, \bibinfo{author}{Z.~Shan}, \bibinfo{author}{J.~Li},
  \bibinfo{author}{X.~Huang}, \bibinfo{author}{L.~Xiao},
  \bibinfo{author}{J.~Sun}, \bibinfo{author}{E.~Ma}, \bibinfo{journal}{Nature}
  \bibinfo{volume}{463} (\bibinfo{year}{2010}) \bibinfo{pages}{335--338}.
\bibitem[{Li and Ma(2009)}]{Li2009b}
\bibinfo{author}{B.~Li}, \bibinfo{author}{E.~Ma}, \bibinfo{journal}{Acta
  Materialia} \bibinfo{volume}{57} (\bibinfo{year}{2009}) \bibinfo{pages}{1734
  -- 1743}.
\bibitem[{Serra and Bacon(1996)}]{Serra1996}
\bibinfo{author}{A.~Serra}, \bibinfo{author}{D.~Bacon}, \bibinfo{journal}{Phil.
  Mag. A} \bibinfo{volume}{73} (\bibinfo{year}{1996})
  \bibinfo{pages}{333--343}.
\bibitem[{Capolungo et~al.(2009)Capolungo, Beyerlein, and
  TomÃ©}]{Capolungo2009}
\bibinfo{author}{L.~Capolungo}, \bibinfo{author}{I.~Beyerlein},
  \bibinfo{author}{C.~TomÃ©}, \bibinfo{journal}{Scripta Mater.}
  \bibinfo{volume}{60} (\bibinfo{year}{2009}) \bibinfo{pages}{32 -- 35}.
\bibitem[{Khater et~al.(2012)Khater, Serra, Pond, and Hirth}]{Khater2012}
\bibinfo{author}{H.~A. Khater}, \bibinfo{author}{A.~Serra},
  \bibinfo{author}{R.~C. Pond}, \bibinfo{author}{J.~P. Hirth},
  \bibinfo{journal}{Acta Materialia} \bibinfo{volume}{60}
  (\bibinfo{year}{2012}) \bibinfo{pages}{2007 -- 2020}.
\bibitem[{Rajabzadeh et~al.(2013{\natexlab{a}})Rajabzadeh, Legros, Combe,
  Mompiou, and Molodov}]{Rajabzadeh2013}
\bibinfo{author}{A.~Rajabzadeh}, \bibinfo{author}{M.~Legros},
  \bibinfo{author}{N.~Combe}, \bibinfo{author}{F.~Mompiou},
  \bibinfo{author}{D.~A. Molodov}, \bibinfo{journal}{Phil. Mag.}
  \bibinfo{volume}{93} (\bibinfo{year}{2013}{\natexlab{a}})
  \bibinfo{pages}{1299--1316}.
\bibitem[{Rajabzadeh et~al.(2013{\natexlab{b}})Rajabzadeh, Mompiou, Legros, and
  Combe}]{Rajabzadeh2013a}
\bibinfo{author}{A.~Rajabzadeh}, \bibinfo{author}{F.~Mompiou},
  \bibinfo{author}{M.~Legros}, \bibinfo{author}{N.~Combe},
  \bibinfo{journal}{Phys. Rev. Lett.} \bibinfo{volume}{110}
  (\bibinfo{year}{2013}{\natexlab{b}}) \bibinfo{pages}{265507}.
\bibitem[{Molodov et~al.(2007)Molodov, Gorkaya, and Gottstein}]{Molodov2007}
\bibinfo{author}{D.~A. Molodov}, \bibinfo{author}{T.~Gorkaya},
  \bibinfo{author}{G.~Gottstein}, \bibinfo{journal}{Mater. Sci. Forum}
  \bibinfo{volume}{558-559} (\bibinfo{year}{2007}) \bibinfo{pages}{927--932}.
\bibitem[{Gorkaya et~al.(2009)Gorkaya, Molodov, and Gottstein}]{Gorkaya2009}
\bibinfo{author}{T.~Gorkaya}, \bibinfo{author}{D.~A. Molodov},
  \bibinfo{author}{G.~Gottstein}, \bibinfo{journal}{Acta Mater.}
  \bibinfo{volume}{57(18)} (\bibinfo{year}{2009}) \bibinfo{pages}{5396--5405}.
\bibitem[{Rupert et~al.(2009)Rupert, Gianola, Gan, and Hemker}]{Rupert2009}
\bibinfo{author}{T.~J. Rupert}, \bibinfo{author}{D.~S. Gianola},
  \bibinfo{author}{Y.~Gan}, \bibinfo{author}{K.~J. Hemker},
  \bibinfo{journal}{Science} \bibinfo{volume}{326} (\bibinfo{year}{2009})
  \bibinfo{pages}{1686--1690}.
\bibitem[{Mompiou et~al.(2009)Mompiou, Caillard, and Legros}]{Mompiou2009}
\bibinfo{author}{F.~Mompiou}, \bibinfo{author}{D.~Caillard},
  \bibinfo{author}{M.~Legros}, \bibinfo{journal}{Acta Materialia}
  \bibinfo{volume}{57} (\bibinfo{year}{2009}) \bibinfo{pages}{2198--2209}.
  \bibinfo{note}{Doi: DOI: 10.1016/j.actamat.2009.01.014}.
\bibitem[{Mompiou et~al.(2011)Mompiou, Legros, and Caillard}]{Mompiou2011}
\bibinfo{author}{F.~Mompiou}, \bibinfo{author}{M.~Legros},
  \bibinfo{author}{D.~Caillard}, \bibinfo{journal}{{Journal of Mater. Science}}
  \bibinfo{volume}{{46}} (\bibinfo{year}{{2011}})
  \bibinfo{pages}{{4308--4313}}.
\bibitem[{Mompiou et~al.(2012)Mompiou, Legros, Sedlmayr, Gianola, Caillard, and
  Kraft}]{Mompiou2012b}
\bibinfo{author}{F.~Mompiou}, \bibinfo{author}{M.~Legros},
  \bibinfo{author}{A.~Sedlmayr}, \bibinfo{author}{D.~Gianola},
  \bibinfo{author}{D.~Caillard}, \bibinfo{author}{O.~Kraft},
  \bibinfo{journal}{Acta Materialia} \bibinfo{volume}{60}
  (\bibinfo{year}{2012}) \bibinfo{pages}{977--983}.
\bibitem[{Eberl et~al.(2006)Eberl, Gianola, and Thompson}]{Eberl2006}
\bibinfo{author}{C.~Eberl}, \bibinfo{author}{D.~Gianola},
  \bibinfo{author}{R.~Thompson}, \bibinfo{title}{{Digital image correlation and
  tracking }}, \bibinfo{howpublished}{MATLAB file exchange,
  http://www.mathworks.com/matlabcentral/fileexchange/12413.},
  \bibinfo{year}{2006}.
\bibitem[{Gianola et~al.(2011)Gianola, Sedlmayr, Monig, Volkert, Major,
  Cyrankowski, Asif, Warren, and Kraft}]{Gianola2011a}
\bibinfo{author}{D.~S. Gianola}, \bibinfo{author}{A.~Sedlmayr},
  \bibinfo{author}{R.~Monig}, \bibinfo{author}{C.~A. Volkert},
  \bibinfo{author}{R.~C. Major}, \bibinfo{author}{E.~Cyrankowski},
  \bibinfo{author}{S.~A.~S. Asif}, \bibinfo{author}{O.~L. Warren},
  \bibinfo{author}{O.~Kraft}, \bibinfo{journal}{Review of Scientific
  Instruments} \bibinfo{volume}{82} (\bibinfo{year}{2011})
  \bibinfo{pages}{063901--12}.
\bibitem[{Richter et~al.(2009)Richter, Hillerich, Gianola, Monig, Kraft, and
  Volkert}]{Richter2009}
\bibinfo{author}{G.~Richter}, \bibinfo{author}{K.~Hillerich},
  \bibinfo{author}{D.~Gianola}, \bibinfo{author}{R.~Monig},
  \bibinfo{author}{O.~Kraft}, \bibinfo{author}{C.~Volkert},
  \bibinfo{journal}{Nano Letters} \bibinfo{volume}{9} (\bibinfo{year}{2009})
  \bibinfo{pages}{3048--3052}.
\bibitem[{Johanns et~al.(2012)Johanns, Sedlmayr, Sudharshan, Mönig, Kraft,
  George, and Pharr}]{Johanns2012}
\bibinfo{author}{K.~Johanns}, \bibinfo{author}{A.~Sedlmayr},
  \bibinfo{author}{P.~Sudharshan}, \bibinfo{author}{R.~Mönig},
  \bibinfo{author}{O.~Kraft}, \bibinfo{author}{E.~George},
  \bibinfo{author}{G.~M. Pharr}, \bibinfo{journal}{Journal of Materials
  Research} \bibinfo{volume}{27} (\bibinfo{year}{2012})
  \bibinfo{pages}{508--520}.
\bibitem[{Cahn and Taylor(2004)}]{Cahn2004}
\bibinfo{author}{J.~W. Cahn}, \bibinfo{author}{J.~E. Taylor},
  \bibinfo{journal}{Acta Mater.} \bibinfo{volume}{52} (\bibinfo{year}{2004})
  \bibinfo{pages}{4887--4898}.
\bibitem[{Serra et~al.(1999)Serra, Bacon, and Pond}]{Serra1999}
\bibinfo{author}{A.~Serra}, \bibinfo{author}{D.~Bacon},
  \bibinfo{author}{R.~Pond}, \bibinfo{journal}{Acta Mat.} \bibinfo{volume}{47}
  (\bibinfo{year}{1999}) \bibinfo{pages}{1425--1439}.
\bibitem[{Braisaz et~al.(1997)Braisaz, Ruterana, Nouet, and Pond}]{Braisaz1997}
\bibinfo{author}{T.~Braisaz}, \bibinfo{author}{P.~Ruterana},
  \bibinfo{author}{G.~Nouet}, \bibinfo{author}{R.~C. Pond},
  \bibinfo{journal}{Phil. Mag. A} \bibinfo{volume}{75} (\bibinfo{year}{1997})
  \bibinfo{pages}{1075--1095}.
\bibitem[{Frank(1965)}]{Frank1965}
\bibinfo{author}{F.~C. Frank}, \bibinfo{journal}{Acta Cryst.}
  \bibinfo{volume}{18} (\bibinfo{year}{1965}) \bibinfo{pages}{862--866}.
\bibitem[{Pond et~al.(1987)Pond, McAuley, and Serra}]{Pond1987}
\bibinfo{author}{R.~C. Pond}, \bibinfo{author}{N.~A. McAuley},
  \bibinfo{author}{W.~A.~T. Serra, A.and~Clarck}, \bibinfo{journal}{Script.
  Mater.} \bibinfo{volume}{21} (\bibinfo{year}{1987})
  \bibinfo{pages}{197--202}.
\bibitem[{Bonnet et~al.(1981)Bonnet, Cousineau, and Warrington}]{Bonnet1981}
\bibinfo{author}{R.~Bonnet}, \bibinfo{author}{E.~Cousineau},
  \bibinfo{author}{D.~H. Warrington}, \bibinfo{journal}{Acta Cryst. A}
  \bibinfo{volume}{37} (\bibinfo{year}{1981}) \bibinfo{pages}{184--189}.
\bibitem[{Sedlmayr et~al.(2012)Sedlmayr, Bitzek, Gianola, Richter, Monig, and
  Kraft}]{Sedlmayr2012}
\bibinfo{author}{A.~Sedlmayr}, \bibinfo{author}{E.~Bitzek},
  \bibinfo{author}{D.~Gianola}, \bibinfo{author}{G.~Richter},
  \bibinfo{author}{R.~Monig}, \bibinfo{author}{O.~Kraft},
  \bibinfo{journal}{Acta Mater.} \bibinfo{volume}{60} (\bibinfo{year}{2012})
  \bibinfo{pages}{3985 -- 3993}.
\bibitem[{Dotsenko(1979)}]{Dotsenko1979}
\bibinfo{author}{V.~I. Dotsenko}, \bibinfo{journal}{physica status solidi (b)}
  \bibinfo{volume}{93} (\bibinfo{year}{1979}) \bibinfo{pages}{11--43}.
\bibitem[{Mompiou(2016)}]{pytrack}
\bibinfo{author}{F.~Mompiou}, \bibinfo{title}{Pytrack},
  \bibinfo{howpublished}{https://github.com/mompiou/PyTrack.git},
  \bibinfo{year}{2016}.
\bibitem[{Gianola et~al.(2006)Gianola, Warner, Molinari, and
  Hemker}]{Gianola2006a}
\bibinfo{author}{D.~S. Gianola}, \bibinfo{author}{D.~H. Warner},
  \bibinfo{author}{J.~F. Molinari}, \bibinfo{author}{K.~J. Hemker},
  \bibinfo{journal}{Scripta Mater.} \bibinfo{volume}{55} (\bibinfo{year}{2006})
  \bibinfo{pages}{649--652}.
\bibitem[{Hirel et~al.(2008)Hirel, Godet, Brochard, Pizzagalli, and
  Beauchamp}]{Hirel2008}
\bibinfo{author}{P.~Hirel}, \bibinfo{author}{J.~Godet},
  \bibinfo{author}{S.~Brochard}, \bibinfo{author}{L.~Pizzagalli},
  \bibinfo{author}{P.~Beauchamp}, \bibinfo{journal}{Phys. Rev. B}
  \bibinfo{volume}{78} (\bibinfo{year}{2008}) \bibinfo{pages}{064109}.
\bibitem[{Hara et~al.(2009)Hara, Izumi, and Sakai}]{Hara2009}
\bibinfo{author}{S.~Hara}, \bibinfo{author}{S.~Izumi},
  \bibinfo{author}{S.~Sakai}, \bibinfo{journal}{Journal of Applied Physics}
  \bibinfo{volume}{106} (\bibinfo{year}{2009}) \bibinfo{pages}{--}.
\bibitem[{Brochard et~al.(2010)Brochard, Hirel, Pizzagalli, and
  Godet}]{Brochard2010}
\bibinfo{author}{S.~Brochard}, \bibinfo{author}{P.~Hirel},
  \bibinfo{author}{L.~Pizzagalli}, \bibinfo{author}{J.~Godet},
  \bibinfo{journal}{Acta Materialia} \bibinfo{volume}{58}
  (\bibinfo{year}{2010}) \bibinfo{pages}{4182 -- 4190}.
\bibitem[{Izumi and Yip(2008)}]{Izumi2008}
\bibinfo{author}{S.~Izumi}, \bibinfo{author}{S.~Yip}, \bibinfo{journal}{Journal
  of Applied Physics} \bibinfo{volume}{104} (\bibinfo{year}{2008}).
\bibitem[{Zhu et~al.(2008)Zhu, Li, Samanta, Leach, and Gall}]{Zhu2008}
\bibinfo{author}{T.~Zhu}, \bibinfo{author}{J.~Li},
  \bibinfo{author}{A.~Samanta}, \bibinfo{author}{A.~Leach},
  \bibinfo{author}{K.~Gall}, \bibinfo{journal}{Phys. Rev. Lett.}
  \bibinfo{volume}{100} (\bibinfo{year}{2008}) \bibinfo{pages}{025502}.
\bibitem[{Okamoto(2000)}]{Okamoto2000}
\bibinfo{editor}{H.~Okamoto} (Ed.), \bibinfo{title}{Phase Diagrams for Binary
  Alloys}, \bibinfo{publisher}{ASM International}, \bibinfo{year}{2000}.
\bibitem[{Serra et~al.(1991)Serra, Pond, and Bacon}]{Serra1991}
\bibinfo{author}{A.~Serra}, \bibinfo{author}{R.~Pond},
  \bibinfo{author}{D.~Bacon}, \bibinfo{journal}{Acta Metallurgica et
  Materialia} \bibinfo{volume}{39} (\bibinfo{year}{1991}) \bibinfo{pages}{1469
  -- 1480}.
\bibitem[{Choi and Auh(1995)}]{Choi1995}
\bibinfo{author}{J.~Choi}, \bibinfo{author}{K.~Auh}, \bibinfo{journal}{{Mater.
  Lett.}} \bibinfo{volume}{{24}} (\bibinfo{year}{{1995}})
  \bibinfo{pages}{{161--165}}.
\bibitem[{Hirth and Lothe(1982)}]{Hirth1982}
\bibinfo{author}{J.~Hirth}, \bibinfo{author}{J.~Lothe}, \bibinfo{title}{Theory
  of Dislocations}, \bibinfo{publisher}{Krieger}, \bibinfo{year}{1982}.
\bibitem[{Coujou et~al.(1990)Coujou, Lours, Roy, Caillard, and
  Clement}]{Coujou1990}
\bibinfo{author}{A.~Coujou}, \bibinfo{author}{P.~Lours},
  \bibinfo{author}{N.~Roy}, \bibinfo{author}{D.~Caillard},
  \bibinfo{author}{N.~Clement}, \bibinfo{journal}{Acta Metallurgica et
  Materialia} \bibinfo{volume}{38} (\bibinfo{year}{1990}) \bibinfo{pages}{825
  -- 837}.

\end{thebibliography}
\bibliographystyle{model1a-num-names}

\end{document}